\documentclass[aps,prb,showpacs,twocolumn]{revtex4}
\usepackage{amsmath,amssymb,amsfonts,mathrsfs}
\usepackage[USenglish]{babel}
\usepackage[dvips]{graphicx}
\usepackage{subfigure}
\usepackage{epsfig}
\usepackage[colorlinks=true,linkcolor=blue,citecolor=red,
bookmarksnumbered=true,bookmarks=true]{hyperref}

\newcommand{\eq}[2]
{
  \begin{equation}
    #1
    \label{#2}
  \end{equation}
}

\newcommand{\eqnn}[1]
{\begin{equation*}
    #1
  \end{equation*}}

\newcommand{\equ}[1]
{Eq.~(\ref{#1})}

\newcommand{\figu}[1]
{fig.~\ref{#1}}

\newcommand{\secu}[1]
{Sec.~\ref{#1}}

\def\bcen{\begin{center}}
\def\ecen{\end{center}}

\def\a{\alpha}          \def\g{\gamma}   \def\d{\delta} 
\def\e{\varepsilon}  \def\z{\zeta}        
        \def\m{\mu}      
          \def\p{\pi}     \def\r{\rho}     \def\s{\sigma}
\def\t{\tau}           
        \def\o{\omega}

\def\GG{{\cal G}}

\def\RRR{\mathbb{R}}  

  \def\ie{\mbox{\it i.e.\ }}
\def\=={\equiv}

\def\qed{\raise1pt\hbox{\vrule height5pt width5pt depth0pt}}
\def\iome{i\omega_n}  
\def\iomn{i\omega_n}
\def\epsk{\epsilon({\bf k})} 
  
\def\cG0{{\cal G}_0} 
\def\cG{{\cal G}}    
 
 \def\intbeta{\int_{0}^{\beta}}
  
\def\up{\uparrow} \def\down{\downarrow} \def\dw{\downarrow}
\def\bra{\langle} \def\ket{\rangle}
\def\ka{{\bf k}}

  \def\Im{\mbox{Im}}
\def\ie{\hbox{\it i.e.\ }} 
\def\ie{\mbox{\it i.e.\ }} \def\=={\equiv}
 
\def\Im{{\rm Im}}  
 \def\ep0{\epsilon_{p}} \def\ed0{\epsilon_{f}}
\def\tpd{V_{fp}}

\begin{document}
\title{A path to poor coherence in heavy fermions from Mott physics and hybridization}
\author{A.~Amaricci$^{1}$, L.~de'~Medici,$^{2}$, G.~Sordi$^{3}$, M.J.~Rozenberg$^{2,4}$,
M.~Capone$^{1,5}$} 
\affiliation{$^1$ CNR-IOM, SISSA, Via Bonomea 265, 34136 Trieste, Italy.}
\affiliation{$^2$ Laboratoire de Physique des Solides, CNRS-UMR8502, Universit\'e Paris-Sud,
Orsay 91405, France.}
\affiliation{$^3$ Theory Group, Institut Laue Langevin, 6 Rue J.~Horowitz, 38042
Grenoble, France.}
\affiliation{$^4$ Departamento de F\'{\i}sica, FCEN, Universidad de Buenos Aires,
Ciudad Universitaria Pab.I, Buenos Aires 1428, Argentina.}
\affiliation{$^5$ Physics Department, University ``Sapienza'', Piazzale A. Moro 2, 00185
Rome, Italy.}

\date{\today}

\begin{abstract}
We investigate the anomalous metal arising by hole doping the Mott insulating state of the 
periodic Anderson model. Using Dynamical Mean-Field Theory we show that, as opposed to the
electron-doped case, in the hole-doped regime the hybridization between localized and
delocalized orbitals leads to the formation of composite quasi-particles reminiscent of the
Zhang-Rice singlets.
We compute the coherence temperature of this state, showing its extremely small value at
low doping. As a consequence the weakly-doped Mott state deviates from the predictions of
Fermi-liquid theory already at small temperatures.
The onset of the Zhang-Rice state and of the consequent poor coherence is due to the
electronic structure in which both localized and itinerant carriers have to be involved in
the formation of the conduction states and to the proximity to the Mott state.
By investigating the magnetic properties of this state, we discuss the relation between the
anomalous metallic properties and the behavior of the magnetic degrees of freedom.
\end{abstract}

\maketitle

%===============================================================================
\section{Introduction}\label{sec0}
The rise of the field of strongly correlated materials revealed a number of unexpected 
intriguing phenomena which can not be explained within the standard theory of solids.
\cite{Ashcroft}
The paradigm of correlation effects is based on the Mott insulating state and the 
Mott-Hubbard metal-insulator transition,\cite{imadaRMP,mott} but a key role is also played
by high-temperature superconductivity in copper oxides \cite{Bednorz86,Anderson87} and
the unconventional superconductivity at the edge of a magnetic phase observed in heavy
fermions.\cite{Flouquet06,CePd2Si2} More recently, the partnership between exotic
superconductivity, strong correlations and magnetism has been strengthened by the
discoveries in the iron-based superconductors,\cite{Hosono08} in the alkali-doped
fullerides\cite{Takabayashi09,Capone09} and possibly also in the molecular conductors based
on aromatic molecules.\cite{kubozono10,Giovannetti11,nomura11}

A common companion of Mott physics and anomalous superconductivity is the deviation 
from the standard Fermi-Liquid (FL) theory in the metallic phase, or non-Fermi-liquid (NFL)
behavior.\cite{nflstewart} The FL theory describes a system of interacting fermions as a 
collection of renormalized non-interacting {\it quasi-particles} which propagate coherently
in the solid.\cite{nozieres97} 
The main qualitative effect of the electron-electron correlations is to enhance the
effective mass and accordingly reduce the coherence of the Fermi gas. This reflects in a
reduction of the coherence temperature, the scale at which the thermal fluctuations destroy
the coherent motion. However in many compounds, most notably heavy fermion materials and
underdoped cuprates, this picture breaks down and the carriers can no longer be described as
long-lived excitations as they acquire a finite lifetime. This behavior directly influences
the transport properties leading to anomalies in the temperature dependence of
the resistivity.

In this paper we present a general mechanism based on Mott physics and multi-band effects 
which leads to a metallic state with an extremely small FL coherence temperature.
Empirically, this system will display a NFL behavior already at exceedingly small
temperatures. The key element is the hybridization between a strongly correlated band and a
weakly interacting band that leads to the formation of hybrid entities. The binding with the
localized $f$-electrons hinders the motion of the carriers leading to a coherence
temperature which is much smaller than the (already renormalized) scale predicted
by FL theory on the basis of mass renormalization.

Our approach is based on the periodic Anderson model (PAM), a widely accepted correlated
electrons model for the description of the heavy fermion physics. In its minimal form the
PAM describes a set of non-dispersive strongly correlated electrons, hybridizing with a band
of conduction electrons. In a general framework the PAM provides a more detailed description
of the electronic configuration of correlated materials with respect to the Hubbard model,
by taking into account the effects of the inclusion of non-correlated bands.
We solved the PAM using dynamical mean-field theory (DMFT),\cite{rmp} one of the most
powerful and reliable tools to study correlated materials.

Following previous studies\cite{sordi07,amaricci08,sordi09} we investigate the model around
the Mott insulating state which takes place for large interactions and {\it odd} integer
total occupation. The doping-driven transition has been thoroughly investigated in
Ref.~\onlinecite{sordi07,sordi09}, and a NFL behavior in the hole-doped side has been
demonstrated in Ref.~\onlinecite{amaricci08}.
Here we extend this work by analyzing the coherence-incoherence crossover which leads to the
NFL behavior and its dependence on doping. We will therefore focus on the scattering 
properties of the system and we will detail their relation with the magnetic degrees of
freedom. Finally we establish a connection between the finite-temperature breakdown of the
FL and the competition between anti-ferromagnetic and ferromagnetic short-ranged
correlations.
 
The manuscript is structured as follows. In \secu{sec1} we introduce the PAM and the
related DMFT equations. In section \secu{sec2} we briefly discuss the doping-driven Mott
transition in the PAM. In \secu{sec3} we present the main results of this work, namely
the strongly incoherent nature of the low temperature metallic state. A phase-diagram of the
model is presented at the end of this section. In \secu{sec4} we study the magnetic
properties of the model. Finally, we present a magnetic phase-diagram of the model which
illustrates how magnetic competition helps stabilizing the incoherent behavior at low
temperature.

\section{Model and theoretical framework}\label{sec1}

\subsection{The Periodic Anderson Model}\label{sec1.1}
The periodic Anderson model describes a set of non-dispersive strongly correlated electrons,
locally hybridizing with a band of conduction electrons. The model Hamiltonian is written in
the following form:
\eq{
\begin{split}
H&= H_0 + H_{fp} + H_{I}\\
H_0&=-\sum_{<ij>\s}t_{ij} p^+_{i\s}p_{j\s}
+\ep0\sum_{i\s}n_{pi\s} +\ed0\sum_{i\s} n_{fi\s}\\
H_{fp}&=\tpd\sum_{i\s}\left(f^+_{i\s}p_{i\s} + p^+_{i\s}f_{i\s}\right) \\
H_I&=U\sum_i\left(n_{fi\uparrow}-\tfrac{1}{2}\right)\left(n_{fi\downarrow}
-\tfrac{1}{2}\right)
\end{split}
}{PAMHam}
The operators $p_{i\s}$ ($p_{i\s}^+$) destroy (create) conduction band electrons with
hopping amplitude $t_{ij}$ and energy $\ep0$.
The operators $f_{i\s}$ ($f_{i\s}^+$) destroy (create) electrons in the non-dispersive
orbital with energy $\ed0$.
The terms proportional to $\tpd$ describe the hybridization between the two species.
The interaction term $H_I$ describes the strong on-site Coulomb repulsion experienced by
$f$-orbital electrons.

The non-interacting lattice Green's function reads:
\eqnn{
\hat{G}_{0\s}^{-1} (\ka,\iome)= 
\left( 
\begin{array}{cc}
\iome+\mu-\ed0 & -\tpd \\
-\tpd & \iome+\mu-\ep0-\epsk
\end{array} 
\right)
}
with $\epsk$ the dispersion of the conduction electrons: $\epsk=
\sum_{<ij>}e^{-i\ka\cdot(\mathbf{r}_i-\mathbf{r}_j)}\, t_{ij}$.
The corresponding interacting Green's function can be expressed by means of the following
matrix Dyson equation:
\eq{
\hat{G}_\s(\ka,\iome)^{-1}=\hat{G}_{0\s}^{-1}(\ka,\iome)-\hat{\Sigma}_\s(\ka,\iome)
}{dyson}
where 
\eq{
\hat{\Sigma}_\s(\ka,\iome)=
\left(
\begin{array}{cc}
\Sigma_{f\s}(\ka,\iome) & 0 \\
0 & 0
\end{array}
\right)
}{sigma}
is the self-energy matrix $\hat{\Sigma}_\s$. The non-interacting nature of the conduction
band is reflected in the existence of only one non-zero self-energy for the $f$-electrons.
Nevertheless, it is useful to define an {\it
effective} self-energy also for the conduction electrons as:
\eq{
\Sigma_{p\s}(\ka,\iome)=\frac{\tpd^2}{\iome+\mu-\ed0-\Sigma_{f\s}(\ka,\iome)}
}{sigmapp}
This function describes the dressing of the $p$-electrons as an effect
of both their hybridization with the correlated $f$-electrons and,
indirectly, of the Hubbard repulsion on the latter. In particular, the
appearance of a finite imaginary part in the zero-frequency limit
signals the breakdown of a FL picture for the conduction electrons.

Since $\Sigma_{p \s}$ arises due to both the hybridization and the
interaction $U$, it is not expected to
vanish in the non-interacting limit $U=0$. 
On the other hand, it is easy to realize that in this limit  the pure hybridization can not
lead to a finite imaginary part of $\Sigma_{p \s}$ at zero frequency, and that any breakdown
of the FL behavior can descend only from correlation effects.

\subsection{DMFT equations}\label{sec1.2}
The PAM has been studied using a large variety of numerical\cite{Grewe88,
Newns87,Schweitzer91,Fazekas87,Pruschke00} and analytical
methods.\cite{canionoce,Gurin01,Gulacsi02,Shiba90} 
To access the non-perturbative regime of the PAM, we investigate the solution of the
model using the DMFT, which has been used to solve this model since its early
stages.\cite{jarrell,marcelo_PAM} 

Within DMFT a lattice model is mapped onto an effective single-impurity problem,
fixed by a self-consistency condition which enforces the equivalence between the two
models as far as the local physics is concerned.\cite{phytoday,rmp} 
The scheme is equivalent to a local approximation on the self-energy, which becomes momentum
independent.

The DMFT equations can be obtained using a quantum cavity method. 
The effective action of the single $f$-orbital impurity problem is obtained integrating out
all lattice degrees of freedom except for a chosen site (labeled conventionally as site
$i=0$) and keeping only the first term in the expansion\cite{metzvol,rmp} in terms of
many-particle Green's functions:
\eq{
\begin{split}
{\rm S_{\rm eff}}=&
-\int_0^{\beta}d\tau \int_0^\beta d\tau' \sum_{\sigma}
f^+_{0\sigma}(\tau)
\GG^{-1}_{0\s}(\tau - \tau')f_{0\sigma}(\tau') \\
& +U\int_0^{\beta} d\tau [n_{f0\up}(\tau)-1/2][n_{f0\down}(\tau)-1/2]
\end{split}
}{Seff}
\begin{figure}
\subfigure[$\,$ doped Kondo Insulator]{\label{fig1.1a}
 \includegraphics[width=0.23\textwidth]{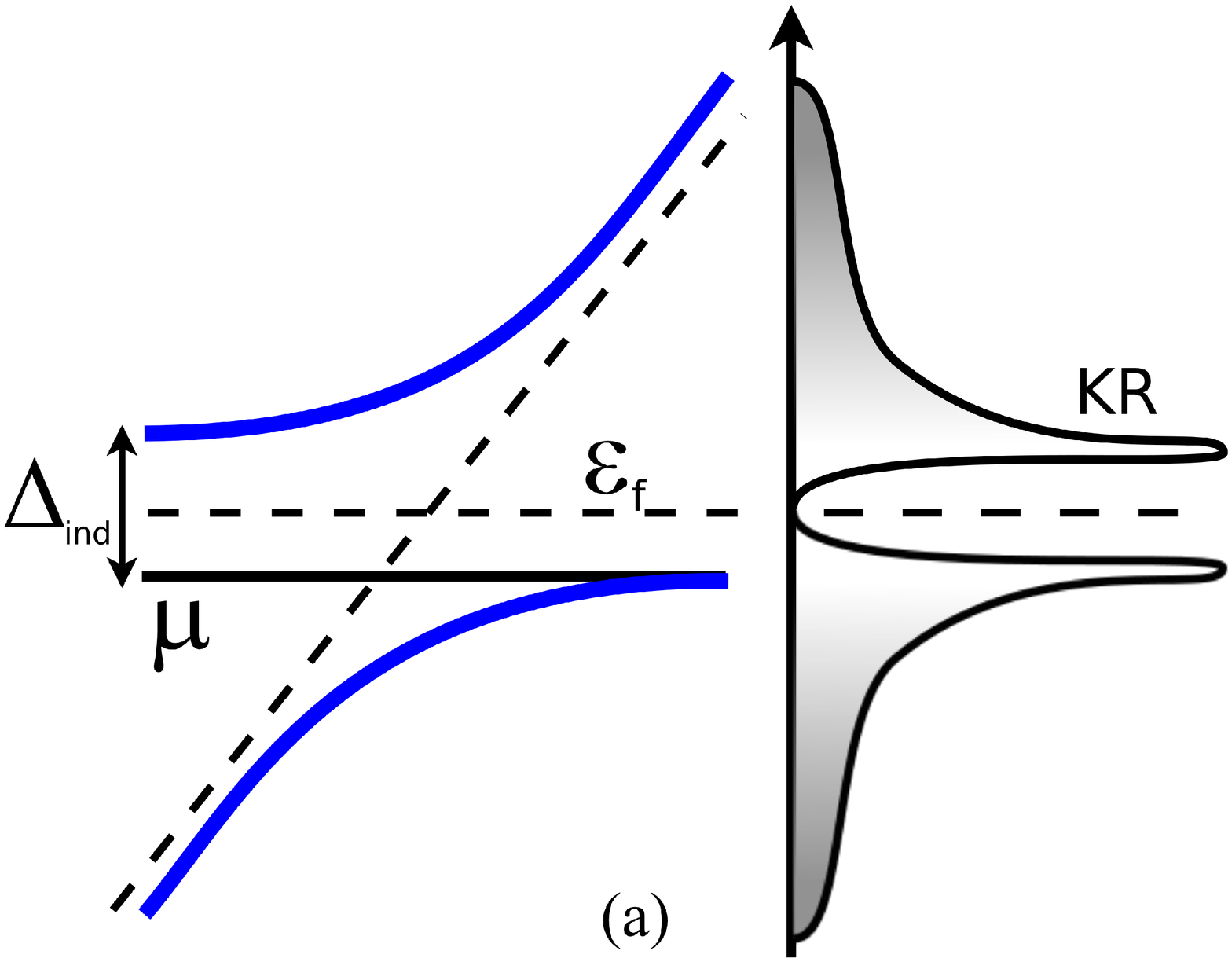}}
 \subfigure[$\,$ doped Mott Insulator]{\label{fig1.1b}
 \includegraphics[width=0.23\textwidth]{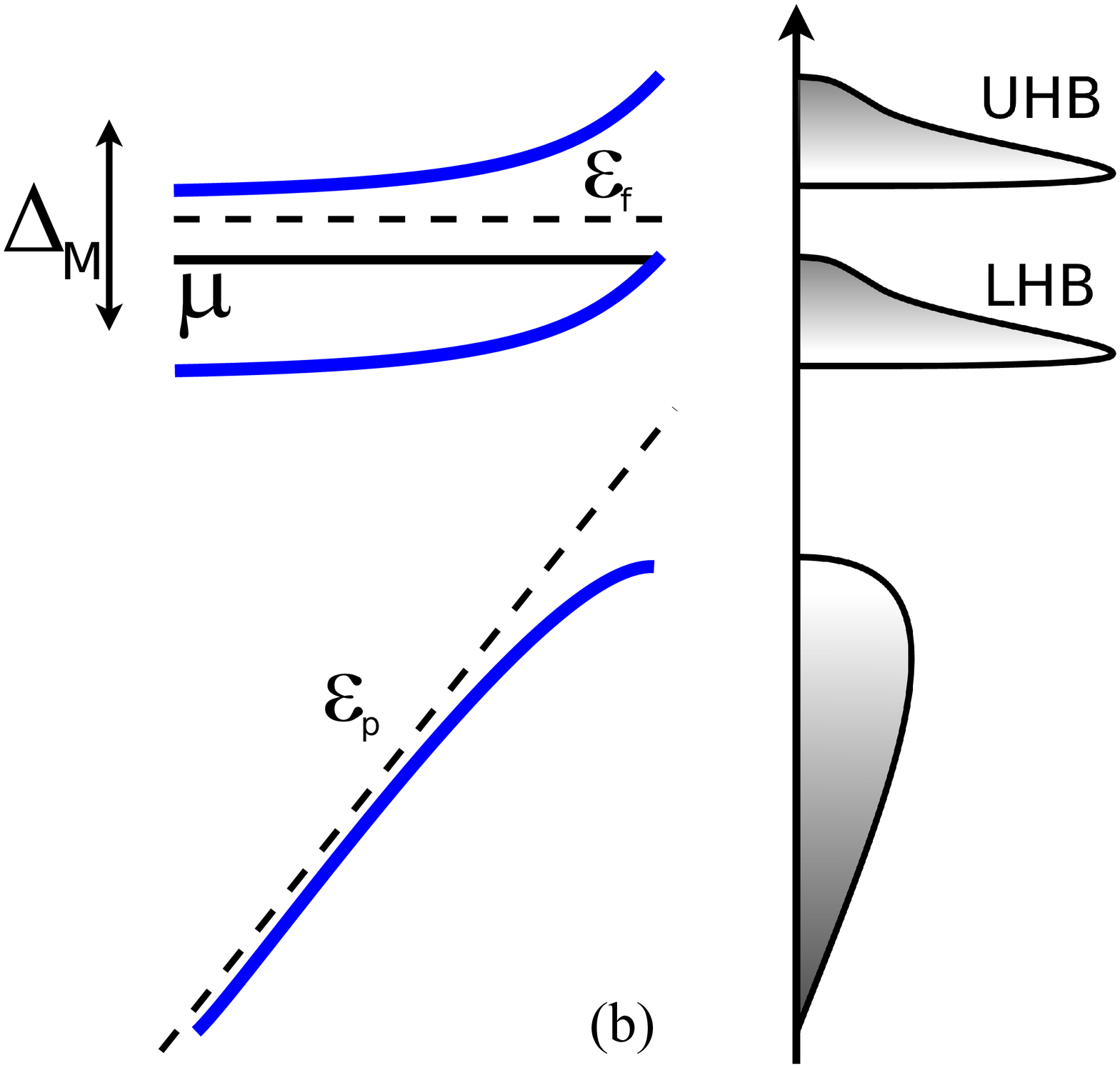}}
\caption{(Color online) Schematic representation of the doped Kondo (a) and doped Mott (b)
insulators, inspired by Ref.~\onlinecite{Coleman07}.
A schematic picture of the band structure is reported in the left side of each figure. 
The hybridized bands are indicated with thick (blue online) lines, dashed lines are
used for the bare bands ($\tpd=U=0$). In the right sides we sketched the corresponding
density of states. A lighter color indicates the more correlated character of the hybridized
bands.
}
\label{fig1.1}
\end{figure}
The action $S_\mathrm{eff}$ is expressed in terms of the local {\it Weiss Field}
$\GG_{0\s}^{-1}(\iome)$, describing the quantum fluctuations at the correlated $f$-orbital.
The Weiss field satisfies a self-consistence equation which depends on the lattice under
consideration. In this work we consider a Bethe lattice with semi-elliptical density of
states of half-bandwidth $D$ (fixing the energy unit of the problem), $D(\varepsilon) =
\frac{2}{\pi D^2}\sqrt{D^2-\varepsilon^2}$.
In this case the self-consistency is particularly simple and reads: 
\eq{
\GG_{0\s}^{-1} (i\omega_n)=
\iome+\m-\ed0 - \frac{\tpd^2}{\iome+\m-\ep0-\frac{D^2}{4}G_{p\s}(\iome)}
}{weiss}
where $G_{p\s}$ is the conduction electron local Green's function.
The functional form of $\GG_{0\s}^{-1}$ mirrors in the DMFT equations the relation between
the two orbitals in the lattice problem. The fluctuations at the $f$-orbital are in
fact composed of two contributions: (a) the on-site quantum fluctuations and (b) indirect
delocalization through conduction band proportional to squared hybridization amplitude. 

The DMFT solution requires therefore to compute the impurity Green's function
\eq{
G^{\rm imp}(\iome)=-i\bra f\, f^+ \ket_{\rm S_{\rm eff}}
}{Gimp}
where the symbol $\bra \, \ket_{\rm S_{\rm eff}}$ indicates the average 
with respect to the effective action (\ref{Seff}). From the knowledge of the impurity
Green's
function it is straightforward to determine the self-energy:
$$
\Sigma(\iome)_\s=\Sigma^{\rm imp}_\s(\iome)=\GG_{0\s}^{-1}(\iome)-{G^{\rm imp}_\s}^{-1}
(\iome)
$$
and finally to evaluate the local Green's function:
$$
G_{p\s}(\iome)=\int
d\varepsilon\frac{D(\varepsilon)}{\iome+\mu-\ep0-\Sigma_{p\s}(\iome)-\varepsilon}
$$
Then a new Weiss field can be computed and the procedure can be iterated until convergence 
is achieved.

The solution of the effective impurity problem, \ie \equ{Gimp}, is the
bottleneck of the DMFT algorithm. In this work we use a combination of numerical techniques:\cite{rmp} Hirsch-Fye Quantum Monte Carlo (QMC) \cite{kotliarRMP,dosSantos03} and Exact Diagonalization (ED) methods, both in the full diagonalization and Lanczos algorithm implementations, at zero\cite{dagottoRMP} and finite temperature.\cite{lucaED}
The ED method is based on a discretization of the effective bath on an adaptive energy grid. In this paper we present full ED calculations in which the bath is described by 7 energy levels and Lanczos calculations with 8 levels. 
The ED calculations have been cross-checked against Density Matrix Renormalization Group, which allows to substantially increase the number of bath levels. 
\cite{dmrgRMP,karenPRB,daniel1}

\section{The hole-doped Mott insulator}\label{sec2}
\begin{figure}
\centering
\includegraphics[width=0.45\textwidth]{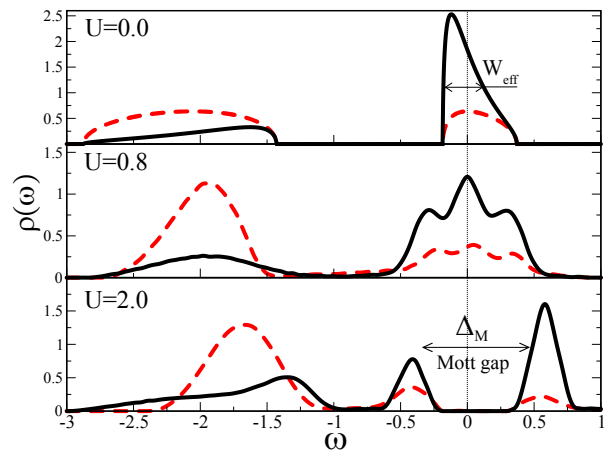}
\caption{(Color online) Evolution of the $f$- (solid line) and $p$-orbital (dashed
line) projections  of the DOS. Data from QMC calculations at $T=0.0125$,
$\tpd=0.9$, $n_{tot}=3$, analytically continued on the real axis using Maximum Entropy
Method \cite{mem}. The figure qualitatively illustrates the Mott metal-insulator transition
driven by correlation in the PAM.}
\label{fig2.1}
\end{figure}

The PAM has been largely investigated in proximity of the Kondo insulator
regime.\cite{schwolff,Fazekas,doniach} 
The Kondo insulator is a band insulator realized at {\it even} integer total filling
($n_\mathrm{tot}=2$).
In this regime the system has two hybridized bands with a central Kondo peak, corresponding
to the resonant scattering of the conduction electrons on the localized moments and split
by an indirect gap $\Delta_\mathrm{ind}$ (see \figu{fig1.1a}).
Upon doping the Kondo resonance remains pinned at the chemical potential,
and the system behaves like a heavy-fermion liquid.

In this work we focus on a different model regime, namely the correlated metal obtained by 
a state with {\it odd} total occupation ($n_\mathrm{tot}=1$ or $3$) and large enough
interaction. In the case of $n_\mathrm{tot}=1$ or $3$, an important role is played by the
ratio $U/\Delta$, where $\Delta=|\ep0-\ed0|$ is the charge-transfer energy, \ie the
separation in energy between the two electron orbitals. Two regimes can be
distinguished:\cite{zsa} 
(a) for $\Delta$ smaller than $U$ the model is in the so-called {\it Charge-Transfer} (CT)
regime, that is expected to capture the properties of intermediate to late transition-metal
oxides. Nevertheless in these systems the non-local hybridizations become important and 
require the introduction of other terms in the Hamiltonian to be properly described. 
(b) For $\Delta$ larger than $U$ the model is in the {\it Mott-Hubbard} (MH) regime, which
models the properties of  early transition-metal oxides and heavy fermion systems, usually
dominated by local physics. In this work we shall focus on this latter model regime and
study the doping of a Mott insulator.

In the simplest sketch of this regime, the non-correlated band has a lower energy than the 
correlated one (which however is dispersive only because of the hybridization with the
itinerant fermions). The latter band is in turn split by the Mott gap (see \figu{fig1.1b}). 
Similarly to the Kondo Insulating regime, a heavy fermion state is obtained upon finite 
doping as soon as the system develops a coherent Kondo resonance signature of the
insulator-metal transition.

\begin{figure}
\centering
\includegraphics[width=0.45\textwidth]{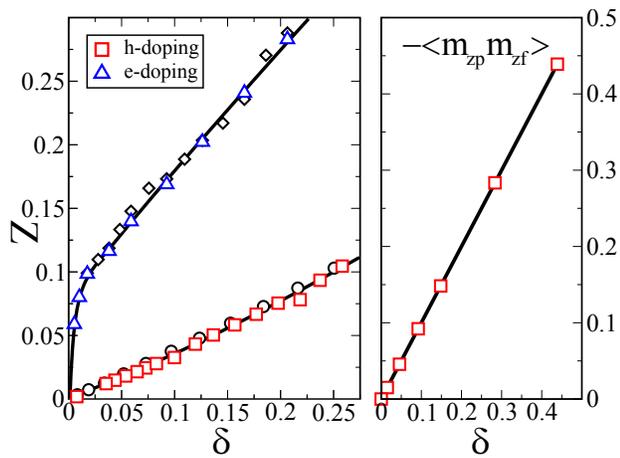}
\caption{(Color online) Left panel: renormalization constant $Z$ as a function of the
doping $\d=|3-n_\mathrm{tot}|$. Data are from Lanczos ED at $T=0$, $U=2$,
$\tpd=0.9$ and increasing size of the effective bath $N_s=8$ (diamonds, circles) and
$10$ (triangles, squares).
Right panel: moment-moment correlation $-\bra m_{zp}\cdot m_{zf}\ket$ as a
function of the doping $\d$ in the hole doped regime. Data are from full ED calculations
(see Appendix~\ref{apx1}) for $T=0.008$ and for the same model parameters.}
\label{fig2.2}
\end{figure}

In the following we shall briefly review the formation of a correlated metallic state by 
hole doping\cite{sordi07,sordi09}. Without loss of generality, we fix the energy of the
correlated orbitals at the Fermi level $\ed0=0$ and $\ep0=-1$, so that $\Delta=1$. For $U=\tpd=0$ the model describes a system with completely filled conduction band $n_p=2$ and half-filled correlated orbitals with $n_f=1$. For finite values of the hybridization the
correlated electrons can move with an effective hopping of the order of $t_{\rm
eff}\simeq\tpd^2/\Delta$, corresponding to the indirect delocalization through the
conduction band (see top panel of \figu{fig2.1}). The hybridization modifies the orbitals
occupation, pushing a substantial amount of the $p$-electron states to the Fermi level, so
that $n_p<2$ and consequently $n_f>1$ and the relevant carriers are hybrid in nature.
However, the $f$- and $p$-character of the model solution can still be used to indicate
the projection onto the correlated and non-correlated orbital respectively. Upon increasing
the interaction strength (see central panel of \figu{fig2.1}) we first observe the formation
of a correlated metallic state. This is characterized by the presence in the density of
states (DOS) of a metallic feature at the Fermi level, flanked by the two precursors of the
Hubbard bands. A Mott insulating state is then obtained further increasing the correlation
$U$. The system opens a spectral gap at the Fermi energy (see bottom panel of \figu{fig2.1})
with a width controlled by the correlation $U$.\cite{sordi09} 
To fix ideas, in the remaining part of this work we shall set the correlation and the hybridization to, respectively, $U=2.0$ and $\tpd=0.9$. This choice of the model parameters corresponds to a Mott insulating state for $n_\mathrm{tot}=3$. Similar results can be obtained for different values of correlation and hybridization.

The Mott insulator can be destabilized in favor of a correlated metallic phase by either
adding or removing electrons (creating holes). 
As first noticed in Ref.~\onlinecite{sordi07} the two transitions have a different
character, ultimately related to the different role played by the non interacting band in
the two cases. Doping with electrons, the extra carriers populate essentially the correlated
orbitals while the p-band remains almost filled and its role is to allow the
delocalization of correlated electrons. In other words, in this regime there are no
multi-band effects and the hybridization plays a minor role. Therefore  the f-electrons
behave essentially as in a single-band Hubbard model with an effective hopping of the order
of $t_\mathrm{eff}$.

\begin{figure}
\centering
 \includegraphics[width=0.45\textwidth]{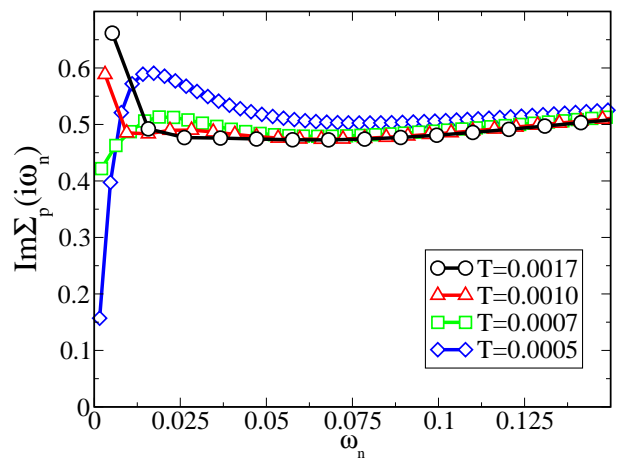}
\caption{(Color online) Evolution of the imaginary part of the conduction electrons
self-energy $\Im\Sigma_p(\iome)$ for increasing temperature. Data are from finite
temperature Lanczos ED with doping $\d=0.1$.
}
\label{fig2.3}
\end{figure}

In the hole doped regime the electronic configuration is substantially different. The holes
are essentially associated to the absence of $p$-electrons, and they tend to bind to the
local moments of the almost half-filled correlated orbitals\cite{sordi09}.
It is already apparent that this state can not be described by a single-band model.
This effect is evident in the behavior of the inter-orbital moment-moment correlation
$$
\bra m_{zp}\cdot m_{zf}\ket
=\bra
(n_{p\up}-n_{p\dw})\cdot(n_{f\up}-n_{f\dw})
\ket
$$ 
reported in \figu{fig2.2}, which shows how the moment 
of the doped p-holes aligns with the moment of the localized f-electron.
The doping-driven metalization appears as the process of delocalizing  a multi-band
``Zhang-Rice-like" singlet state, formed by an itinerant hole bound to a localized spin,
similar to that proposed in the framework of the high-T$_c$ superconductors.\cite{zr} 
The low-energy properties of this metallic state can not be straightforwardly
interpreted in terms of  a single-band Hubbard model,\cite{sordi07,sordi09} and it leads
to remarkable properties.

A first partial indication of the anomalous nature of this state comes from an evaluation 
of the quasi-particle weight
$Z=[1-\partial\Im\Sigma_f(i\o)/\partial\omega]_{|_{\omega\rightarrow0}}^{ -1}$, 
which measures the degree of metallicity of a system, being zero for a Mott insulator and
one for a non-interacting metal. The results (see \figu{fig2.2}) show that $Z$ is
substantially smaller for the hole-doped than for the electron-doped case, already signaling
that the Zhang-Rice liquid is a poorer metal than a standard correlated metal. In the
following we will show that the difference goes well beyond the quasi-particle
renormalization.

\begin{figure}
\centering
\includegraphics[width=0.45\textwidth]{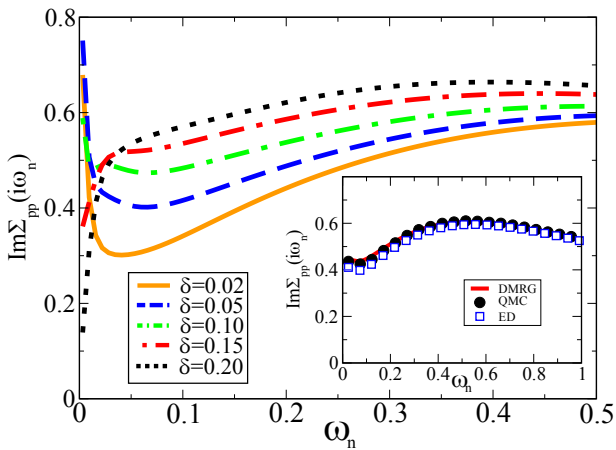}
\caption{(Color online) Main panel: imaginary part of the conduction electrons self-energy
$\Im\Sigma_p(\iome)$ for increasing value of the hole doping and $T=0.001$. Data are from
Lanczos ED calculations.
Inset: Comparison of the $\Im\Sigma_p(\iome)$ behavior from different numerical methods for $\d=0.05$. The other model parameters are the same as in the main panel.  The QMC and
full ED calculations are performed at $T=0.008$. DMRG is a $T=0$ calculation performed with
a cluster of $N_{\rm s}=30$ sites and plotted down to the position of the lowest energy
pole.}
\label{fig2.4}
\end{figure}

\section{Thermal breakdown of the Fermi-liquid}\label{sec3}
Fermi-liquid theory is the standard paradigm for metallic systems and describes correlated
Fermi systems as a collection of non-interacting renormalized quasi-particles. DMFT studies
of various correlated models have shown that even very close to the Mott
transition the correlated metallic state is typically a Fermi liquid with a reduced 
effective hopping proportional to the quasi-particle weight $Z$. This scale also controls the
coherence temperature above which the coherent motion of the carriers is destroyed by
thermal fluctuations.

In this section we will show that  the correlated metallic state of the PAM in
the weakly hole-doped regime turns out to be very fragile with respect to small
temperatures. More precisely, our system will be a Fermi liquid only below an extremely small coherence temperature which, for small doping, can be substantially smaller than 
the renormalized Fermi energy controlled by $Z$. Therefore the corresponding metallic 
state can not be described in terms of long-lived quasi-particles but is rather a liquid
of short-lived singlet-like electronic excitations.

\begin{figure}
\centering
\includegraphics[width=0.45\textwidth]{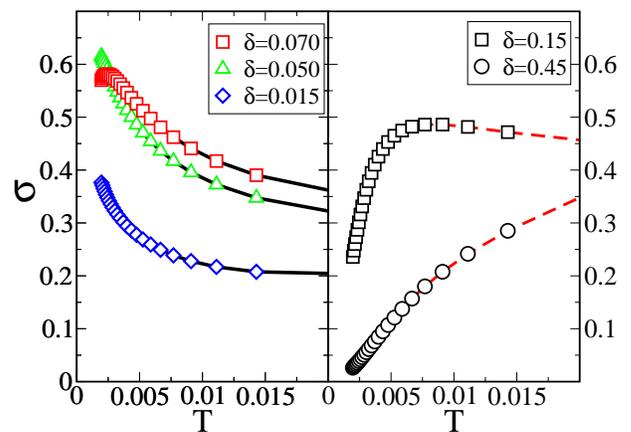}
\caption{
$\s=\Im\Sigma_p(\iome\rightarrow0)$ as a function of the temperature
for different values of the doping. The data shown are from full ED calculations.
}
\label{fig2.5}
\end{figure}

To substantiate this discussion we study the evolution of the imaginary part of the
conduction electron self-energy $\Im\Sigma_p(\iome)$.
The results of our calculations for $\delta =0.1$ are presented in \figu{fig2.3}.
A Fermi liquid state corresponds to a linear behavior of $\Im\Sigma_p(\iome)$ at low 
frequency, observed only at the lowest investigated temperature, $T=0.0005$. 
When we increase $T$ at values of the order of $T=0.0007$, two orders of magnitude smaller
than the {\it renormalized} Fermi energy, the conduction-electron self-energy does not
vanish in the $\omega \to 0$ limit, signaling a departure from the Fermi-liquid paradigm. 
Further increasing the temperature leads to an enhancement of this anomaly.

In \figu{fig2.4} we follow the evolution of $\Im\Sigma_p(\iome)$ for increasing 
doping at $T=0.001$. For small doping we have a clear NFL increase at small frequency which
survives up to $\delta \simeq 0.16$. For larger doping the system is not strongly sensitive
to the Mott-Hubbard physics and the standard Fermi-liquid behavior is restored around
$\delta = 0.2$.

The violation of the Fermi-liquid paradigm can be summarized by the temperature
dependence of $\s(T) = \Im\Sigma_p(\iome\rightarrow0)$, reported in \figu{fig2.5}. 
This quantity is related to the scattering rate of the carriers. 
In a metallic regime $\s(T)$ is expected to vanish at low temperature. While for large doping (right panel) $\s(T)$ vanishes as $T \to 0$ (even if for $\delta = 0.15$ some anomaly is observed at intermediate temperature), the small-doping data clearly confirm the NFL behavior down to very small temperature, even if, strictly at $T=0$ the vanishing $\s$ would be recovered. 

\begin{figure}
\centering
\includegraphics[width=0.45\textwidth]{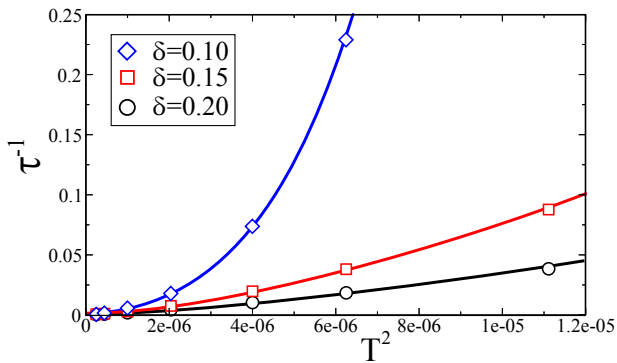}
\caption{(Color online) Scaling of the inverse life-time $\tau^{-1}$ as a function of
$T^2$ in the small temperature limit. Data are from Lanczos ED calculations. Lines
are guide to the eye.
}
\label{fig2.6}
\end{figure}

Finally, \figu{fig2.6} depicts  the inverse life-time $\t^{-1}=Z_p \s$ of the
doped carriers, where $Z_p^{-1}=1-\Im\Sigma_p(i\o_1)/\p T$. In a Fermi liquid $\t^{-1}$ 
grows as $T^2\sim \o^2$ at low temperature. Our calculations for small doping show a decay
faster than $T^2$ which strengthens the picture of an incoherent metallic state. Once again,
a Fermi-liquid behavior is established only at extremely low temperatures if the doping is
small, while the large-doping data recover the standard behavior.

The increasing scattering rate as a function of decreasing temperature is usually
associated to scattering with impurities\cite{nflstewart}. In this spirit, in the following 
we will interpret our results as the scattering of the carriers with fluctuating local
moments. This effect can be understood as the results of the competition between the
aforementioned tendency to form local Zhang-Rice-singlets, driven by the hole-doping, and
the incoherent nature of the scatterer provided by the $f$-electron local moments, driven by
Mott physics. 
At large doping the increased number of available holes of $p$-type helps the formation
of a many-body coherent state without breaking the local binding with $f$-moment.
This arguments will be substantiated by the calculations that we report in the following 
sections.

\subsection{The coherence temperature}\label{sec3.1}
The analysis of the self-energy and of the carriers lifetime clearly shows the existence of a small doping-dependent energy scale associated with the appearance of an incoherent metal.
We expect this scale to influence also other observables, like the local spin
susceptibility:
$$
\chi_\mathrm{loc}(T)=\intbeta\bra S_{zf}(\t)\cdot S_{zf}(0)\ket  d\t
$$

This quantity describes the response to a {\it local} magnetic field and easily discriminates 
between a Fermi-liquid, in which the zero-temperature limit is a constant (Pauli susceptibility), 
and a paramagnetic Mott insulator in which it diverges like $1/T$ (Curie behavior).

\begin{figure}%[!ht]
\centering
\includegraphics[width=0.45\textwidth]{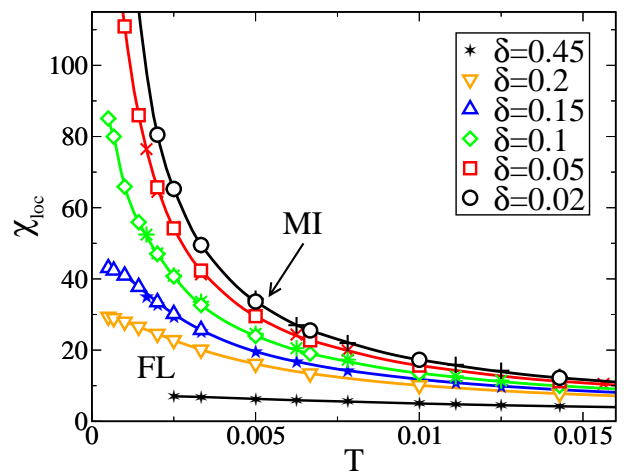}
\caption{(Color online) Local spin susceptibility $\chi_\mathrm{loc}$ as a function of the
temperature and increasing value of the hole doping. Data are from Lanczos ED (open symbols)
and full ED (pluses, crosses and stars symbols) calculations.
}
\label{fig2.8}
\end{figure}

The results are reported in \figu{fig2.8}. In the Mott insulating state ($\d=0$) the magnetic moments 
of the localized $f$-electrons essentially behave as free spins, we thus obtain the typical Curie 
behavior with a $1/T$ dependence for the spin susceptibility.
The slightly hole-doped regime does not show the behavior of a standard metal, namely $\chi_\mathrm{loc}$ 
keeps on increasing down to the lowest investigated temperature $T\simeq
10^{-3}D$ without any sign of saturation. The enhancement of the spin susceptibility signals
the presence of unquenched local moments and can be associated to protracted screening
effect.\cite{tahv1} Only for larger doping, the susceptibility saturates to large constant
value at very low temperature.

The presence of enhanced low-$T$ spin susceptibility coexisting with a (bad) metallic
behavior substantiates the idea that the hole-doped system can be regarded as formed by 
nearly free (incoherent) moments, and an underlying metallic host formed by the doped holes
which are prevented from coherently delocalize by local coupling to $f$-moments.
% binding into singlet state. 
This interpretation leads us to estimate the coherence temperature $T_\mathrm{coh}$ from
\eq{
\chi^{-1}_\mathrm{loc}(T)\propto T+T_\mathrm{coh}
}{chiFIT}

\begin{figure}%[!ht]
\centering
\includegraphics[width=0.45\textwidth]{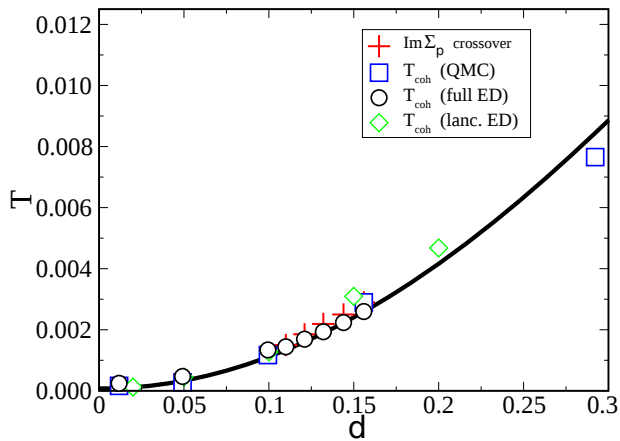}
\caption{(Color online) Coherence temperature scale $T_\mathrm{coh}$ as extrapolated from
the inverse local spin susceptibility $\chi^{-1}_\mathrm{loc}(T)$. The extrapolations from
different numerical methods are found to be in satisfactory agreement. 
}
\label{fig2.7}
\end{figure}

We plot the resulting values, obtained with different numerical methods, in \figu{fig2.7}.
In the same plot we report the crossover points estimated from the temperature evolution of
the imaginary part of the self-energy $\Sigma_p$ (red crosses). 
The good agreement of these points with the extrapolated data validates the physical
interpretation of the coherence temperature.  It is unfortunately very
hard to identify the functional form of the coherence temperature due
to the smallness of the scale involved and the numerical uncertainties.
However,  the data are compatible with an exponential behavior of the form
$T_\mathrm{coh}\simeq Be^{-A/\d}$, which has been obtained within the $1/N$ approximation in
the infinite-$U$ Kondo limit\cite{burdin}. 

\begin{figure}%[!ht]
\centering
\includegraphics[width=0.5\textwidth]{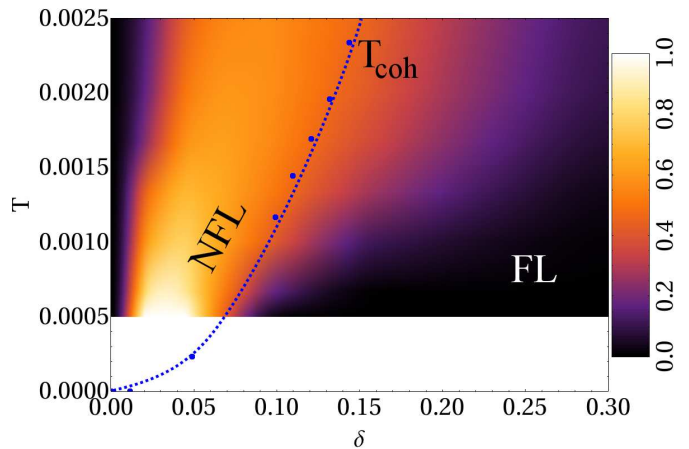}
\caption{(Color online)
Phase-diagram of the PAM near the Mott insulating state as a function of temperature and
hole-doping. The diagram is obtained from $\Im\Sigma_p(\iome\rightarrow0)$. The dotted line
indicates the crossover temperature scale $T_\mathrm{coh}$. 
}
\label{fig2.9}
\end{figure}

The phase diagram in the doping-temperature plane, presented in \figu{fig2.9}, can help us
to summarize the scenario emerging from our calculations. The diagram reveals the character
of the DMFT solution in proximity of the Mott insulating state through the behavior of the $\Im\Sigma_p(\iome\rightarrow0)$. 
Using finite temperature Lanczos ED method we investigated a smaller temperature scale with
respect to that studied in Ref.~\onlinecite{amaricci08}.
A large value of the $\Im\Sigma_p(\iome\rightarrow0)$ testifies a NFL behavior and the results
clearly show that the highly incoherent state emerges from the Mott state and occupies a
sizable region of the phase diagram. 
The NFL region is separated from the coherent metal by a crossover taking 
place at $T_\mathrm{coh}$ defined above, which therefore confirms its meaning as the
temperature in which the metal loses coherence. 

% % % % % % % % % % % % 
% % % % % % % % % % % % 
% % % % % % % % % % % % 
% % % % % % % % % % % % 
% % % % % % % % % % % % 

\section{Magnetic properties}\label{sec4}
\subsection{External magnetic field}\label{sec4.1}
We have shown that hole-doping the Mott insulating phase of the periodic Anderson
model leads to peculiar charge carriers, so that the motion of the created $p$-holes occurs
through the formation of Zhang-Rice singlets, in which the spins of the conduction electrons
are anti-ferromagnetically correlated with the localized spins. 
As a consequence, we expect that a magnetic field can have important and surprising effects
on this phase, showing a further
difference with respect to a standard Fermi liquid.

In the model regime investigated in this work, the main source of magnetism comes from the
$f$-electrons. The conduction band is almost completely filled, so that the magnetization
of the few singly occupied orbitals, favored by hole doping, is not expected to contribute
significantly to the magnetic properties of the system.

\begin{figure}%[!ht]
\centering
\includegraphics[width=0.45\textwidth]{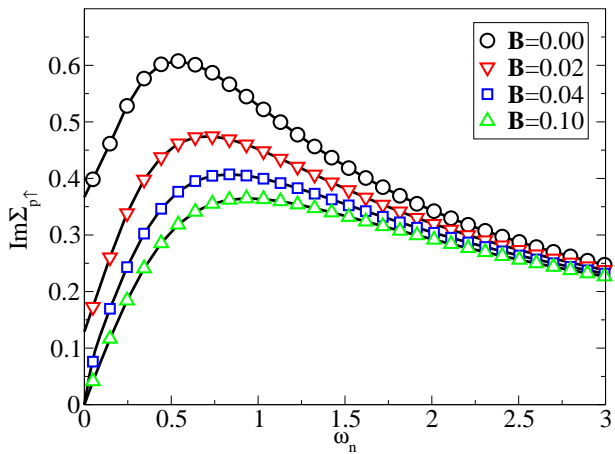}
\caption{(Color online) Imaginary part of the majority-spin $p$-electron
self-energy for increasing external magnetic field $\mathbf{B}$.
The data are from QMC solution at $T=0.016$ and $\d=0.05$. 
}
\label{fig3.1}
\end{figure}

Nevertheless, conduction band electrons can be indirectly affected by the
magnetic polarization of the $f$-orbital moments, through their local binding.
To illustrate this point, we show in \figu{fig3.1} the evolution of the low energy
part of $\Im\Sigma_p(\iome)$ as a function of a uniform magnetic field
$\mathbf{B}$.
Apparently the NFL state turns into a normal metallic state by the action of an external 
magnetic field. It is however worth noting that the Fermi liquid is recovered for
$B\simeq0.05D$, a huge value if compared with experimentally accessible fields.
This large value is a direct consequence of the large (order one) value of $\tpd$, chosen to
emphasize the hybridization effects and their role in the conduction properties of the
model. Smaller and more realistic values of this parameter are expected to reduce the 
critical field by reducing the charge fluctuations at correlated $f$-orbitals.

The crossover to a Fermi liquid state driven by external magnetic field is not surprising in
light of our analysis. Upon increasing the magnetic field a larger and larger number of
local $f$-moments are polarized. When the moments are aligned, the $p$-holes can move
essentially freely in the ferromagnetic background without breaking the singlet state with
the localized spins.
Therefore the source of scattering disappears and the metallic state recovers the
Fermi-liquid coherence.
In other words the polarization of $f$-orbital local moments allows the conduction 
electron cloud to dynamically screen the correlated electrons local moments, dramatically
increasing the coherence scale of the system.

\begin{figure}%[!ht]
\centering
\includegraphics[width=0.45\textwidth]{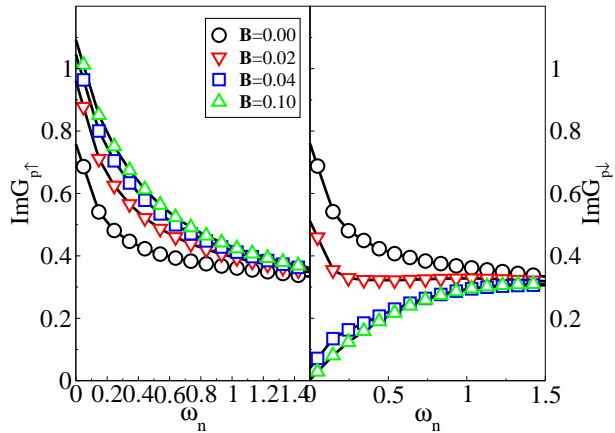}
\caption{Imaginary part of conduction band electron Green's function $\Im G_{p\s}(\iome)$
for $T=0.016$, $\d=0.05$ and increasing strength of external magnetic field. Data from QMC
calculations.
}
\label{fig3.2}
\end{figure}

The coherent motion of the doped carriers with the opposite spin of the localized momenta 
(majority spin) should then be balanced by the insulating nature of the minority spins
carriers.
This effect is illustrated in the \figu{fig3.2}. In this figure we show the
behavior of both spin species conduction electrons Green's function for the
same strengths of the external magnetic field as used in \figu{fig3.1}.
Left panel shows the increasing metalization of the majority spin charge carriers, whereas
in the right panel we show how minority spins are driven towards an insulating state
by increasing magnetic field.

\subsection{Anti-ferromagnetic ordering}\label{sec4.2}
At low temperature, we expect the development of anti-ferromagnetic (AFM) correlations
a result of super-exchange between neighboring $f$-electrons assisted by the hybridization 
with $p$-orbital states. In this section we investigate the onset of an AFM long-range
ordered state and its effect on the coherence scale using the extension of the DMFT
equations to long-range order detailed in Appendix \ref{apx2}.

To begin with we report in \figu{fig3.3} the staggered magnetization $m_{AF}=
1/N\sum_i (-1)^{i} \bra n_{fi\up}-n_{fi\dw}\ket$ as a function of the temperature for
various doping. The transition appears to be of second order in the whole
doping region. The N\'eel temperature $T_N$, extracted from a power-law fit of the data, is
maximum at zero doping and decreases by adding holes, as in the single-band Hubbard model. 
\footnote{The nature of the transition makes the precise determination of the
doping value at which the ordering temperature vanishes numerically hard.
Nevertheless, the data available at smaller doping concentrations suggest the AFM region 
to be bounded by $\delta=0.1$ at zero temperature}

% , reported in the inset of the figure. 
% As in the single-band Hubbard model $T_N$ is maximum at zero doping and decreases by
% adding holes. 
%This is because the energy gain in developing AFM correlations
%between neighboring orbitals is not reduced by local singlet formation. 

The onset of an AFM ordering of the local $f$-moments reinstates the Fermi liquid properties
in the tiny hole-doped regime. This effect is illustrated in the left panel of
\figu{fig3.4}, where we present the evolution of the imaginary part of $\Sigma_{p\s}(\iome)$
from the paramagnetic NFL phase to the AFM ordered phase. The large finite intercept present
in NFL phase is driven to zero in the AFM ordered phase. Nevertheless, the metallic
character of the solution is preserved across the transition, as illustrated in the right
panel of the same figure by comparing the imaginary parts of the conduction electrons
Green's functions in the two phases.
The ordering of the local moments in a N\'eel state, allows the doped charge carriers to
form coherent electronic waves (with doubled wave-vector) and to get delocalized. However, 
as mentioned above, the AFM state is only stable in a small window of doping and the NFL 
remains stable for a wide range of parameters.

\begin{figure}
\centering
\includegraphics[width=0.45\textwidth]{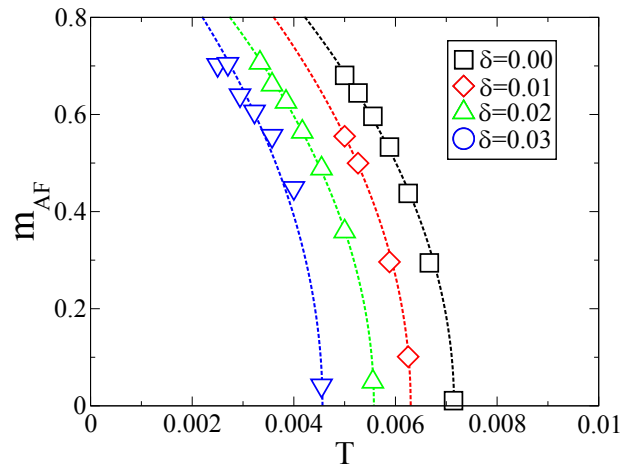}
\caption{(Color online) Main panel: Staggered magnetization $m_{AF}$
as a function of temperature and increasing value of hole-doping. 
The data are from full ED calculations.
}
\label{fig3.3}
\end{figure}

% \subsection{Magnetic stability: $\mathbf{B}$-T phase diagram}\label{sec4.3}
\subsection{Magnetic stability}\label{sec4.3}
The common wisdom about systems of concentrated impurities described by the PAM 
is that long-range magnetic ordering is likely to set in, especially if the metallic
state is weakened by correlations as in our case. 

Our results for the AFM state suggest instead a remarkable stability of the incoherent 
metallic state as the long-range order is confined to low temperature and small doping
concentration. In this section we discuss the physical origin of this surprising result.

At small doping near the Mott insulating state neighboring $f$-orbital electrons develop AFM  
correlations as a results of super-exchange. These processes
are of the fourth order in the hybridization with a leading energy scale of the order:
\cite{Fazekas} $J_{SE} \propto {W_{\rm eff}}^2/U\sim \tpd^4/\Delta^2 U$

\begin{figure}%[!ht]
\centering
\includegraphics[width=.45\textwidth]{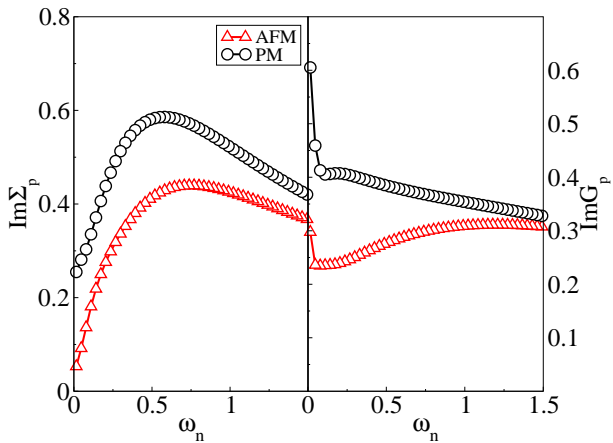}
\caption{(Color online) Conduction electrons self-energy $\Im\Sigma_{p\s}(\iome)$
(left panel) and Green's function $\Im{G}_p(\iome)$ (right panel). Data from full ED
calculations for $\d=0.01$ and $T=0.005$.
}
\label{fig3.4}
\end{figure}

On the other hand  it is easy to realize that at large doping  {\it ferromagnetic}
correlations are expected because of the fact that the doped carriers are locked in 
singlets with the localized $f$-spins. In an AFM or disordered background, the motion of the
$p$-holes requires to break the singlet and it is therefore strongly inhibited, leading to
the lack of coherence that we discussed at length. Moreover it leaves a local moment
unscreened, increasing the fluctuations in the local magnetization. 
Conversely, a ferromagnetic alignment of the localized spins allows for an unperturbed
delocalization of the carriers, with a mechanism which is closely reminiscent of the
double-exchange,\cite{Zener51,Anderson55,deGennes60} where the coupling between
conduction electrons and localized spins is given by the ferromagnetic Hund's coupling.

Therefore, upon increasing the doping the tendency to form AFM ordering is contrasted by
the increased relevance of the kinetic energy and eventually it becomes more favorable to
sacrifice the gain in super-exchange energy in order to gain the kinetic energy associated
to the ferromagnetic background. This leads, most importantly, to an intermediate region
between the two regimes in which the local magnetization is strongly fluctuating. 

From this discussion it is natural to associate the fluctuations of the local 
magnetization to the scattering mechanism that leads to the poor coherence. To test this
idea we study the response of the system in the AFM ordered metallic phase to the
application of a uniform magnetic field $\mathbf{B}$ which will clearly favor the
ferromagnetic tendency.

\begin{figure}%[!ht]>

\centering
\includegraphics[width=0.5\textwidth]{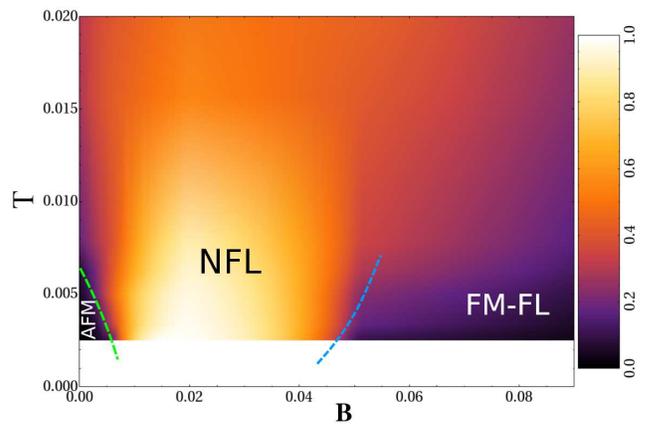}
\caption{(Color online) Intensity plot of $\Im\Sigma_p(\omega\to0)$ as a
function of external magnetic field $\mathbf{B}$ and $T$ at fixed doping $\d = 0.01$. For
visualization, the data have been normalized to $\max\{\Im\Sigma_p(\omega_n)\}$ at
each ($\mathbf{B}$, $T$). Dashed lines are drawn to better visualize the crossover regions
in the phase-diagram.
}
\label{fig3.5}
\end{figure}

The results are summarized in the phase diagram of \figu{fig3.5}, determined again using
$\Im\Sigma_p(\iome\rightarrow0)$. Details of the calculations are given in Appendix
\ref{apx2}.
Doping is fixed to $\d=0.01$, safely into the AFM ordered region in the limit $\mathbf{B}\rightarrow0$.

The phase-diagram shows that the AFM order survives the effects of the external magnetic
field up to small strengths ($B\simeq0.01$). In this region the solution keeps the
coherent metallic character enforced by the long-range magnetic ordering. Nevertheless, for
larger values of the magnetic field the system is driven to an incoherent state with
finite-temperature NFL behavior, as indicated by the increased scattering (light
color). In this regime the large applied field tends to magnetically polarize the AFM
ordered $f$-moments, producing their strong frustration and ultimately leading to the
formation of an incoherent magnetic background for the motion of the doped carriers.
Further increasing the strength of the magnetic field triggers the formation of a
ferromagnetic ordering of the $f$-moments and a fully polarized (coherent) metallic state
(right dark coloured area). 

The most striking observation is that the present diagram faithfully mirrors the diagram 
as a function of doping, clearly suggesting that the evolution of the conduction properties
as a function of doping is associated to the transition from the AFM state to the
ferromagnetic regime and that the poorly coherent metal establishes precisely in the
intermediate region, dominated by the local spin fluctuations which appear as the source of
the scattering mechanism which opposes to the coherent motion of the holes.

\section{Conclusions and perspectives}\label{conclusion}
In this work we presented a detailed dynamical mean-field theory study of the properties of
the unconventional metallic state obtained by doping with holes the Mott insulator in the
periodic Anderson model. We discuss in details the non-Fermi liquid behavior of the system
and the mechanism that is behind the suppression of the coherence scale.

In this regime the holes have mainly $p$-character, but they tend to bind to the 
correlated $f$-electrons  to form a Zhang-Rice-like singlet state. The formation of this
composite object leads to a highly  incoherent metallic state which deviates from a standard
Fermi-liquid above a coherence temperature which decreases very rapidly upon reducing doping, and it is much smaller than the  effective Fermi energy that one could estimate from the degree of correlation of the system, \ie $T_\mathrm{coh}\ll ZD$.
 
We characterize this anomalous behavior by studying the scattering properties of the carriers and by computing the inverse lifetime and local spin susceptibility, which allow us to quantitatively estimate the coherence temperature characterizing the breakdown of the standard Fermi liquid and to describe the onset of an incoherent metal with finite lifetime.
 
The highly incoherent metal is unstable towards anti-ferromagnetic ordering only at very 
small doping, while at large doping ferromagnetic correlations develop and
favour a regular metallic behavior supported by a mechanism which reminds the double-exchange
physics. The intermediate region, where the motion of the holes is not coherent, is
therefore dominated by large fluctuations of the $f$-spins, which provide the scattering
channel responsible of the finite lifetime of the carriers.

The relation between magnetic fluctuations and the breakdown of the standard FL scenario is
emphasized by observing that an external uniform magnetic field, which obviously destroys
AFM ordering favoring a ferromagnetic alignment, mirrors the effect of doping and leads 
again to a wide region of high incoherence between the two magnetically ordered states.

We emphasize that the path to poor coherence discussed in this paper only depends on two
general features of  strongly correlated materials, namely the Mott physics which leads
to the localization of carriers and multi-orbital physics necessary to the local singlet 
formation. In this light, we expect that the mechanism outlined here can be a rather
general source of violation of Fermi liquid paradigm and incoherent behavior, and it can
be relevant for example to heavy fermions, but also, with some important differences 
related to the d-wave symmetry of the Zhang-Rice singlets, to the cuprate superconductors.

Finally, a natural question to address is to what extent our findings can be considered 
the local portrait of the presence of a quantum critical point, hidden by the absence of
spatial fluctuations. Indeed, the existence of a quantum critical point in the PAM, although
in a different model regime, has already been pointed in Ref.~\onlinecite{civdeleo}, using
cluster extension of the DMFT. The development of our work along this direction, in order to
clarify the fate of the small coherence scale in presence of short-range
spatial fluctuations, is left for future research.

\begin{acknowledgements}
A.A., G.S. and M.R. thank M.~Gabay, D.J.~Garcia, E.~Miranda for the many useful discussions
and suggestions. A.A. is also grateful to V.~Dobrosavljevi\'c. 
A.A. acknowledges support from the ESRT Marie-Curie program during part of this work. 
L.dM. acknowledges support from the Agence Nationale de la Recherche (ANR-09-RPDOC-019-01)
and the RTRA ``Triangle de la Physique".
A.A. and M.C. are financed by European Research Council under FP7/ERC Starting Independent 
Research Grant ``SUPERBAD" (Grant Agreement n. 240524)
\end{acknowledgements}

\appendix
\section{Two-orbital effective impurity model}\label{apx1}
The calculation of physical quantities internal to the local ``$pf$-dimer``, such as the
moment-moment correlation function $\bra m_{zp}\cdot m_{zf}\ket$
can be performed within single-site DMFT using an alternative formulation of the effective
impurity problem in which the local p-orbital is not integrated out in the construction of
the effective action. Thus the original lattice system is reduced to the problem of a
single dimer embedded in an electronic bath. The corresponding effective
action has a $2\times 2$ matrix structure in the orbital space and reads:
\eqnn{
\begin{split}
\hat{S}'_{\rm eff}= &-\int_0^{\beta}d\tau \int_0^\beta d\tau'
\sum_{\sigma}\psi^+_{0\s}(\tau) \hat{\cal G}_0^{-1}(\tau -\tau')
\psi_{0\s}(\tau')\\
+& \, U\intbeta d\tau [n_{f0\up}(\tau)-1/2][n_{f0\down}(\tau)-1/2]
\end{split}
}
The Weiss Field $\hat{\cal G}_0^{-1}(\iome)$ describes the local quantum
fluctuations at the tagged dimer. The Bethe lattice self-consistency becomes
\eqnn{
\hat{\cal G}_0^{-1}(\iomn) =
\left(\begin{array}{cc}
\iomn +\mu -\ep0 -\frac{D^2}{4} \,G_p(\iome) & -\tpd \\
-\tpd & \iomn  +\mu -\ed0
\end{array} \right)
}

The DMFT algorithm for the two-orbital representation proceeds as in the standard casse.
The effective two-orbital impurity problem is solved to determine the impurity Green's
functions:
$$
G^{\mathrm imp}_\a(\iome)=-i\bra \a\, \a^+ \ket_{\hat{S'}_{\rm eff}}
$$
with $\a=p,\, f$.
Next, the conduction electron self-energy $\Sigma_p$ can be determined using the Dyson 
equation and used to evaluate the local Green's function $G_p$ which is necessary to update
the local Weiss field. The whole algorithm is iterated until convergence
is reached.

\section{Long range order}\label{apx2}
The DMFT equations can be extended to describe phases with long range
magnetic ordering\cite{rmp}.
Here we derive the equations for the anti-ferromagnetic order in the two-orbital effective
problem, considering also the effect of a uniform magnetic field. 
Similar equations can be derived for the single-orbital effective model.

On a bipartite lattice crystal as our Bethe lattice, we can define two sub-lattices $A$ 
and $B$, such that nearest-neighbor hopping always connects one $A$-site with a $B$-site.
Then we can introduce a four-component spinor with orbital and sub-lattice indices so that
the bare lattice propagator takes the form:
$$
\hat{G}_{0\ka\s}^{-1}=
\begin{pmatrix}
 \a_{A} & -\epsk & -\tpd & 0 \\
 -\epsk & \a_{B} & 0 & -\tpd \\
 -\tpd & 0 & \iome-\ed0+\m_{A} & 0 \\
 0 & -\tpd & 0 & \iome-\ed0+\m_{B}
\end{pmatrix}
$$
with $\a_{s}=\iome-\ep0+\m_{s}$ and $s=A,B$. The corresponding Green's functions
are obtained via the Dyson equation with the diagonal self-energy matrix with
components $\{0,0,\Sigma_{A\s},\Sigma_{B\s}\}$.
The $p$-electrons local Green's functions, required to close the DMFT equations, now read:
\eqnn{
% \begin{split}
G_{pA\s}(\ka,\iome)
% &=\sum_\ka\frac
% {\a_{B\s}-\tpd^2/\g_B}
% {\frac{\tpd^4}{\g_A\g_B}-
% \tpd^2 \left( \frac{\a_{A\s}}{\g_B} + \frac{\a_{B\s}}{\g_A} \right)+
% \a_{A\s}\a_{B\s}-\epsk^2 }\cr
=\sum_\ka\frac{\z_{B\s}}{\z_{A\s}\z_{B\s} - \epsk^2}
% \end{split}
}
where:
\eqnn{
\begin{split}
\z_{s\s}& =\a_{s}-\frac{\tpd^2}{\g_{s\s}}\cr
\g_{s\s}&=\iome-\ed0+\m_{s}-\Sigma_{s\s}(\iome)
\end{split}
}

In the case of anti-ferromagnetic ordering it is not necessary to take explicitly into account 
both sublattices. Observing that: 
$$
\Sigma_{A\s}(\iome)=\Sigma_{B-\s}(\iome)=\Sigma_\s(\iome)
$$
and thus:
\eq{
\z_{A\s}=\z_{B-\s}=\z_{\s}
}{B1}
we can eliminate one of the two sublattices and recover a $2\times 2$ formalism with a 
Weiss field given by
$$
\hat{\cal G}_{0\s}^{-1}(\iomn) =
\left(\begin{array}{cc}
\a_\s -\frac{D^2}{4} \,G_{p-\s}(\iome) & -\tpd \\
-\tpd & \iomn  +\mu_\s -\ed0
\end{array} \right)
$$

The local conduction electron Green's function $G_{p\s}(\iome)$ can be expressed in terms of
the following Hilbert transform:
\eqnn{
G_{p\s}(\iome)=\z_{-\s} \int_\RRR d\e \frac{\r_0(\e)}{\z_\s\z_{-\s}-\e^2}
}
which closes the set of DMFT equations.

In presence of a uniform magnetic field $\mathbf{B}$ in the ordered phase of the
system, the symmetry relation \equ{B1} between the two sublattices does not hold.
Therefore the DMFT solution requires to explicitly consider the two sublattices and the
self-consistency equations for the four components of the Weiss field  $\GG_{0\s s}(\iome)$ 
($s=A,B$ and $\overline{s}=B,A$) read
\eqnn{
\GG_{0\s s}^{-1}=\iome+\overline{\m}_{s\s}-\ed0 -
\frac{\tpd^2}{\iome+\overline{\m}_{s\s}-\ep0-\frac{D^2}{4}G_{p\s\overline{s}}(\iome)}
}
where the coupling to the magnetic field $\mathbf{B}$ has been included in a
redefinition of the chemical potential $\overline{\mu}_{s\,\s}=\m_{s\s}+\s\mathbf{B}/2$.
This means that at each iteration we need to solve two impurity models, one for 
each sub-lattice and that the solution of one sub-lattice will determine the Weiss field
for the other.

\bibliography{bibliografia}
\end{document}